\documentclass[%
 reprint,
 amsmath,amssymb,
 aps,
prc
]{revtex4-2}

\usepackage{graphicx}% Include figure files
\usepackage{dcolumn}% Align table columns on decimal point
\usepackage{bm}% bold math
\usepackage{hyperref}% add hypertext capabilities
\usepackage{color}
\usepackage{ulem}
\usepackage{multirow}
\usepackage{amsmath}
\usepackage{longtable}
\usepackage{CJK}
\usepackage{blindtext}
\usepackage{array}
\usepackage{tabularx}
\newcolumntype{M}[1]{>{\centering\arraybackslash}m{#1}}

\begin{document}

\begin{CJK*}{UTF8}{gbsn}

%\preprint{APS/123-QED}

\title{Average pairing correlation properties and effective pairing  residual interactions}

\author{Meng-Hock Koh (辜明福)}
 \email{kmhock@utm.my}
 %\homepage{https://people.utm.my/kohmenghock/}
 \affiliation{Department of Physics, Faculty of Science, Universiti Teknologi Malaysia, 
        81310 Johor Bahru, Johor, Malaysia.}%Lines break automatically or can be forced with \\

\author{P. Quentin}%
 \email{quentin@cenbg.in2p3.fr}
\affiliation{LP2I, UMR 5797, Universit\'{e} de Bordeaux, CNRS, F-33170, Gradignan, France
}%

\author{L. Bonneau}%
 \email{bonneau@lp2ib.in2p3.fr}
\affiliation{LP2I, UMR 5797, Universit\'{e} de Bordeaux, CNRS, F-33170, Gradignan, France
}%

\date{\today}% It is always \today, today,

\begin{abstract}

This paper describes a method to determine the intensities of effective pairing residual interactions, 
extending what has been done for the seniority force model [Phys. Rev. C 110, 024311 (2024)]. 
It has been tested in Hartree-Fock plus BCS calculations using residual pairing zero-range interactions. 
The average pair condensation energy is the key quantity connecting the determination 
of constant pairing matrix elements to the estimation of delta interaction intensities. 
From individually fitted delta pairing strengths of $28$ well and rigidly deformed nuclei 
whose proton number $Z$ ranges from $50$ to $82$ evaluated at the ground-state, 
we have determined average interaction intensities. 
They reproduce equally well the data on MoI as what is obtained within the seniority force ansatz 
with a r.m.s. deviation of about $2 \: \hbar^2 \mbox{MeV}^{-1}$. 
This approach provides a non-ambiguous way to determine reasonably well the strengths of pairing interactions 
at the ground-state of well deformed nuclei. 
It allows to perform, with some reasonable level of confidence, 
calculations for other nuclei in the corresponding nuclear region as well as 
beyond their ground states in particular to assess deformation properties as, 
e.g., to evaluate fission barriers or spectral properties of quasi-particle states.

\end{abstract}

\maketitle
\end{CJK*}

\section{Introduction \label{sec: introduction}}
In two recent papers \cite{Koh_Philippe_2024,Hao_et_al_2024}, 
a simple yet efficient method has been proposed to determine average pairing matrix elements 
used in calculations within the Hartree-Fock (with self-consistent blocking when necessary) 
plus pairing correlations  
within the standard simple seniority force ansatz (HFBCS-S). 
In many practical uses of such a phenomenological  approach, 
the contact with relevant experimental data is made through some fit of pairing gaps related 
somehow to odd-even mass differences. 
This has yielded various formulae as, e.g., those of Refs. \cite{Jensen1984,Madland1988} providing some 
simple analytical expressions for their dependence on the nucleonic numbers $N$ and $Z$. 
It is important to remark that  making a fit on large enough nuclear regions is akin 
of making a semiclassical average. 
This gives a hint on how a contact could be established between microscopic calculations 
as in our  HFBCS-S approach and such average gaps.

As shown in Ref. \cite{Jennings1975} performing some lengthy self-consistent semi-classical calculations 
(for instance within the Wigner-Kirkwood approximation) could be replaced in practice, up to a very good approximation, 
through a standard (energy) smoothing \textit{\`{a} la} Strutinsky 
\cite{Strutinsky_1966,Strutinsky_1967,Strutinsky_1968,FunnyHill} 
of our microscopic solutions. 
This has provided us, in particular, with average single-particule (sp) level densities $\tilde{\rho}_q (e)$ 
associated with the HF (or BCS) canonical basis, where the subscript $q$ specifies the neutron (n) or proton (p) charge state.

As briefly recalled below, combining such level densities with the fitted smooth estimates of pairing gaps 
for a given nucleus $\Delta_q (N,Z)$, one obtains within the gap equation 
of the so-called uniform-gap method \cite{FunnyHill,RingSchuck}
pairing matrix elements $V_q$ averaged over an interval of sp energies (of span $2\omega$) 
for sp states active in the BCS treatment, centered around the Fermi energies of each charge state. 
These matrix elements are then used within  HFBCS-S calculations. 
This approach can be iteratively rendered consistent for a given nucleus and 
possibly so performed for a limited sample of nuclei yielding through some fit or 
extrapolation to matrix elements valid for other neighbouring nuclear species.

It is worth noting that \textit{a priori} the use of such an approach 
should be reserved exclusively to the description of ground states of well and rigidly deformed nuclei. 
The limitation to nuclear ground states is trivial in view of the origin of the data under consideration. 
The restriction to the above defined deformed nuclei is motivated as follows: 
\begin{itemize}
    \item in doing so we minimize quantal shape fluctuations, maximizing the relevance 
        of a single mean-field wavefunction description;
    \item we reject nuclei too close to magic nuclei, avoiding thus BCS solutions 
        corresponding to weak pairing correlation regimes where the BCS approximation 
        is known to be rather poor \cite{Zheng_1992};

    \item we allow to describe with a good approximation nuclear ground states 
        in the lab frame from intrinsic states within the Bohr-Mottelson unified model 
        (within the pure rotor approximation when describing in particular the moments of inertia) \cite{Bohr_Mottelson_book}.
\end{itemize}
This considerably limits the range of a sound use of the whole approach.

By all means this treatment of pairing correlations could hardly be considered as really new. 
However, in Ref.~\cite{Koh_Philippe_2024} two important features not taken into account in most previous similar works, 
including ours, have been considered.
First, P. M\"{o}ller and J.R. Nix have noted in Ref. \cite{Moller1992} 
that the choice of a sample restricted to well and rigidly deformed nuclei, legitimate as we have seen, 
introduces nevertheless a systematic bias when one is trying to reproduce pairing gap formulae 
as those of Refs. \cite{Jensen1984,Madland1988}. 
Indeed the retained nuclei do correspond to regions of less--dense--than--average sp spectra. 
They have thus proposed in Ref. \cite{Moller1992} gap formulae suitably corrected to take this into account. 
We have adopted their formulae.

The second improvement which we have implemented in our approach 
concerns a systematic distortion of the proton sp spectra upon using the local Slater approximation 
for Coulomb exchange terms \cite{Slater} as achieved in most Hartree-Fock or Hartree-Fock-Bogoliubov calculations 
(not to mention those using an Energy Density Formalism where this is always the case by necessity). 
As noted years ago \cite{Titin}, confirmed later \cite{Skalski} and further systematically studied \cite{Bloas}, 
using this approximation leads to a systematic and significant spurious enhancement 
of the proton sp level density near the Fermi surface. 
This has been approximately corrected for.

\r{A}. Bohr, B.R. Mottelson and D. Pines \cite{BMP}  have singled out 
two nuclear spectroscopic properties affected strongly by 
nuclear pairing which are possibly accessible to mean-field approaches. 
They must therefore be used to estimate the parameters of any such phenomenological approach of pairing correlations. 
They are:
\begin{itemize}
    \item Moments of inertia of well and rigidly deformed even-even nuclei 
    as close to adiabatic rotational mode as possible, that is through the energy of the first $2^+$ state, 
    allowing thus to perform an Adiabatic Time-Dependent Hartree-Fock-Bogboliubov description 
    within an Inglis-Belyaev approach \cite{Inglis_1956,Belyaev}
    corrected approximately \cite{Libert_moi_factor} 
    by the so-called Thouless-Valatin self-consistency corrections \cite{Yuldashbaeva_1999}.

    \item Odd-even mass differences in well and rigidly deformed nuclei, 
    through a three point mass difference formula centered on odd-neutron or odd-proton nuclei \cite{Dobaczewski_2001,Duguet_2001}.
    
\end{itemize}

In Ref. \cite{Hafiza2019} two separate fits of the average pairing matrix elements $V_q$ 
within a sample of well and rigidly deformed nuclei in (and near) the region of rare-earth nuclei 
on either the moments of inertia (of 24 even-even nuclei), 
or the odd-even mass differences (concerning 17 odd-$N$ and 14 odd-$Z$ nuclei), 
using the SIII Skyrme interaction \cite{SIII} have resulted in very similar results.
The calculations of Refs. \cite{Koh_Philippe_2024,Hao_et_al_2024} have provided values of $V_q$ 
of the same quality as those of Ref. \cite{Hafiza2019} \textit{a priori} better suited to reproduce the data. 
This has provided a clear test of the relevance of the whole approach.

Nevertheless, the above, making use of 
the crude seniority force ansatz, clearly suffered from two limitations. 
First, it does not provide direct information on pairing residual interactions 
since it implies an extraneous knowledge of wavefunctions that is somewhat averaged over the considered  
sp states retained in the BCS treatment 
(providing no state dependence of the matrix elements). 
Second, it is restricted \text{a priori} to ground states of well and rigidly deformed nuclei, 
because only there could we warrant, as already discussed, a relatively safe handling of the data.

To extend the study of pairing correlation properties to other nuclei as well as to other states 
(belonging to nuclei with odd number(s) of nucleons, isomeric states, potential energy surfaces as fission barriers etc.), 
one needs to define
the intensity parameters of an interaction by making use of
the information on pairing correlations gained from where
it has been possible to safely relate to data.

This is the goal of this paper to discuss the validity of this determination of pairing residual interactions.
It is organized as follows. 
In Section~\ref{sec:Section II} the method and results of the seminal approaches of Refs. \cite{Koh_Philippe_2024,Hao_et_al_2024} 
making use of the seniority force ansatz will be briefly summarized for self-containedness. 
The method proposed to transfer information from such an approach to treatments 
involving pairing residual interactions will be described in Section~\ref{sec:Section III}. 
Section~\ref{sec:Section IV} will present the results of a first test of the above, 
namely by calculating the moments of inertia of 23 well and rigidly deformed even-even nuclei 
in (and near) the rare-earth region using density-independent (or volume) delta $|T_z| = 1$ (for neutrons and protons) interactions. Finally, some conclusions and perspectives will be sketched in Section~\ref{sec:Section V}.

\section{Average pairing matrix elements: Overview of the method and results
\label{sec:Section II}}

In this Section, we will briefly review for the sake of consistency the method proposed in Ref.~\cite{Koh_Philippe_2024} 
and the results of its tests discussed in Refs.~\cite{Koh_Philippe_2024,Hao_et_al_2024} in the seniority force case.

For the ground state of a chosen nucleus, given the eigenvectors \{$|\phi_i \rangle$\}
and eigenvalues \{$e_i$\} of a one-body Hamiltonian including a mean field potential 
(\textit{a priori} self-consistent or not), 
one computes average sp level densities for protons and neutrons $\tilde{\rho}_q$ averaged 
 \textit{\`{a} la} Strutinsky (see, e.g., Ref.\cite{NPA207})
\begin{equation}
    \tilde{\rho}_q (e) = \frac{1}{\gamma} \int_{-\infty}^{\infty} \rho_q (e') \: f\Big(\frac{e - e'}{\gamma}\Big) de'
\end{equation}
where $\rho_q (e)$ is the exact sp energy density "function" (better said distribution) and with
\begin{equation}
    f(x) = \frac{1}{\sqrt{\pi}} \sum_{\mu = 0}^M \frac{(-1)^\mu}{2^{2\mu} \mu!} \frac{d^{2\mu}}{dx^{2\mu}} \mbox{exp}(-x^2)
\end{equation}
where in our case $M = 2$ and $\gamma = 1.2 \times 41 \times A^{-1/3}$ MeV adopting the usual
sp energy scaling of Ref.~\cite{Moszkowski}.

The averaged Fermi energies $\tilde{\lambda}_q$ are such that
\begin{equation}
    N_q = \int_{-\infty}^{\tilde{\lambda}_q} \tilde{\rho}_q (e) \: de
\end{equation}
with $N_q$ standing for $N$ or $Z$.

We consider the phenomenological average gap values $\Delta_q$ of M\"{o}ller and Nix \cite{Moller1992} 
suitably corrected for protons when treating in self-consistent calculations 
making use of the Slater approximation for the Coulomb exchange terms (see Section~\ref{sec: introduction}). 
%The constant matrix elements $V_q$ to be used 
The absolute value $V_q$ of the constant matrix elements to be used
in the seniority force pairing treatment 
are then obtained upon solving the gap equation
\begin{equation}
    \frac{1}{V_q} = \int_{\tilde{\lambda}_q - \omega}^{\tilde{\lambda}_q + \omega} 
            \frac{\tilde{\rho}_q (e)}{\sqrt{(e - \tilde{\lambda}_q)^2+\Delta_q^2}} \: de
    \label{eq: uniform gap equation}
\end{equation}
with $\omega = 6$ MeV.

Two tests of the validity of this method have been performed. 
They consisted in checking how well some pieces of data were reproduced by Hartree-Fock--plus--BCS calculations 
within the seniority force approximation with matrix elements determined as described above. 
In all these tests, some Skyrme effective force parametrisations were chosen 
for the particle-hole sector of the nucleon-nucleon interaction 
(namely the SIII \cite{SIII}, the SkM* \cite{SkMs} and the SLy4 \cite{SLy4} parameter sets).
Some numerical details of the calculations yielding our HFBCS solutions are briefly summarized 
in Appendix~\ref{Appendix:HFBCS numerical details}.

The first test \cite{Koh_Philippe_2024} was concerned with the moments of inertia of about 20 ground states 
of well and rigidy deformed even-even nuclei in two regions of the nuclear chart: 
the rare-earth region (actually adding some nuclei slightly above it) and the actinide region. 
The data were reproduced with r.m.s. error values displayed in Table~\ref{tab: rms moi seniority} 
for each of the three above specified parametrisations. 
They correspond to relative errors ranging (depending on the retained interactions) 
from 5 to 8 percent in the rare-earth region, and 5 to 7 percent for the actinides. 
A useful point of comparison is provided by the results obtained in Ref.~\cite{Hafiza2019}
upon performing a direct fit of the average pairing matrix elements as described in Section~\ref{sec: introduction}. 
They concerned a sample of rare-earth nuclei about twice as large as compared 
to what has been considered in Ref.~\cite{Koh_Philippe_2024} and they used only the SIII parametrisation. 
Their r.m.s error value (1.75 $\hbar^2$ MeV$^{-1}$) is found indeed very close with ours. 
It is thus reasonable to infer that the magnitude of these deviations from data 
reflect mostly on the average the quality of the s.p. spectra of the interaction in use.

\begin{table}[h]
    \begin{ruledtabular}
        \caption{The r.m.s. deviation of moment of inertia in $\hbar^2 \cdot \mbox{MeV}^{-1}$ unit 
                from experimental data obtained in the work of Ref.~\cite{Koh_Philippe_2024} and \cite{Hafiza2019}}
        \label{tab: rms moi seniority}
        \begin{tabular}{*{5}c}
            Region&     SIII&   SkM*&   SLy4&   \cite{Hafiza2019}   \\
            \hline
            Rare-earth&  1.77&     2.71&      2.94&    1.75    \\
            Actinide&      4.58&   3.39&       3.50&     --   \\
        \end{tabular}
    \end{ruledtabular}
\end{table}

The second test dealt with the odd-even mass difference measured through 3-points mass differences 
centered on odd neutron or proton ground states. 
They were limited to rare-earth nuclei (16 odd-$N$ and 13 odd-$Z$) and to the Skyrme SIII interaction. 
The r.m.s. errors were found to be 78 keV for odd-neutrons and 182 keV for odd-protons. 
Here too, the results which have been obtained through the direct fit of Ref.~\cite{Hafiza2019}
for mostly the same sample of nuclei, 
are very similar (r.m.s. errors of 87 and 182 keV respectively) 
leading to the same conclusion which has been drawn above
about the origin of the discrepancy with the data.

\section{Determination of the residual pairing interactions
\label{sec:Section III}}
As presented in Section~\ref{sec: introduction}, 
we will describe now how the information on pairing correlations gained in the 
seniority case for relevant nuclear ground states
may be used to determine the intensities of pairing residual interactions.

The quantity which we propose to consider for performing such a transfer 
is the (absolute value) of the pair condensation energy which is defined in the BCS approximation, 
for each charge state (dropping the $q$ subscript for a while) as a sum on all Cooper pairs
\begin{equation}
    \tilde{E}^{pc} = - \sum_i \tilde{k}_i \: \tilde{\Delta}_i
\end{equation}
where %the sum runs over all Cooper pairs,
the superscript tilde refers to quantities which are averaged \textit{\`{a} la} 
Strutinsky and $k_i$ is defined in terms of occupation $v_i$ and vacancy $u_i$ amplitudes as
\begin{equation}
    k_i = u_i v_i.    
\end{equation}

As well known the gaps $\Delta_i$ are defined in terms of the pairing residual interaction $v_{res}$ as
\begin{equation}
    \Delta_i =  - \sum_j \langle j \bar{j} | v_{res}^{as} | i \bar{i} \rangle \: k_j
\end{equation}
with $v_{res}^{as} = (1 - P^M P^\sigma P^\tau) v_{res}$
where $P^M, P^\sigma, P^\tau$ are the coordinate (Majorana), spin and isospin exchange operators.

Given a sp spectrum, one computes as described in Section~\ref{sec:Section II}  
an average sp level density $\tilde{\rho}$ and the corresponding averaged Fermi energy $\tilde{\lambda}$.
The averaged quantities $\tilde{k}_i$ are then defined in terms of the Strutinsky smoothing function $f(x)$ as
\begin{equation}
    \tilde{k}_i = \frac{1}{\gamma} \int_{-\infty}^{\infty} 
            f \Big( \frac{e - e_i}{\gamma}\Big) \: k(e,\tilde{\lambda},\Delta_i) de
\end{equation}
with
\begin{equation}
    k(e,\lambda,\Delta) = \frac{\Delta}{2 \sqrt{(e - \lambda)^2 + \Delta^2}}.
\end{equation}

The absolute value of the average pair condensation for each charge state $q$ 
(see Appendix~\ref{Appendix:seniority treatment})
is given in the seniority case by
\begin{equation}
    \tilde{E}_q^{pc} (sen) = E_q^{pc} (sen) = \frac{\Delta_q^2}{V_q}.
\end{equation}

\section{Assessment of our method to a volume delta pairing residual interaction
\label{sec:Section IV}}

In this Section, we consider a particular kind of pairing residual interaction, namely volume delta interaction 
(for each charge states $q$):
\begin{equation}
    v_{res}^q (\Vec{r}_1 - \Vec{r}_2) = W_q \: \delta (\Vec{r}_1 - \Vec{r}_2).
\end{equation}

We aim to assess the relevance of the intensity parameters $W_q$
of such an interaction by comparing the resulting moments of inertia calculated as described in Section~\ref{sec:Section II} 
with available data for a relevant sample of deformed nuclei.

As well known, the quality of the reproduction of these moments is for a large part
contingent upon the quality of the sp spectra in use, that is of the mean-field from where
they are obtained and ultimately upon the quality of the particle-hole interaction. 
The latter is here of the usual Skyrme  analytical form. 
Our present goal is not to assess the quality of the sp spectra but of the process 
allowing to transfer adequately the information gained in the seniority force case 
in order to define a pairing residual interaction. 
This implies that such a test should make use of a parametrisation of the Skyrme
interaction which is known from previous studies to provide reasonable sp spectra in the considered region.

\begin{figure}
    \centering
    \includegraphics[width=\linewidth]{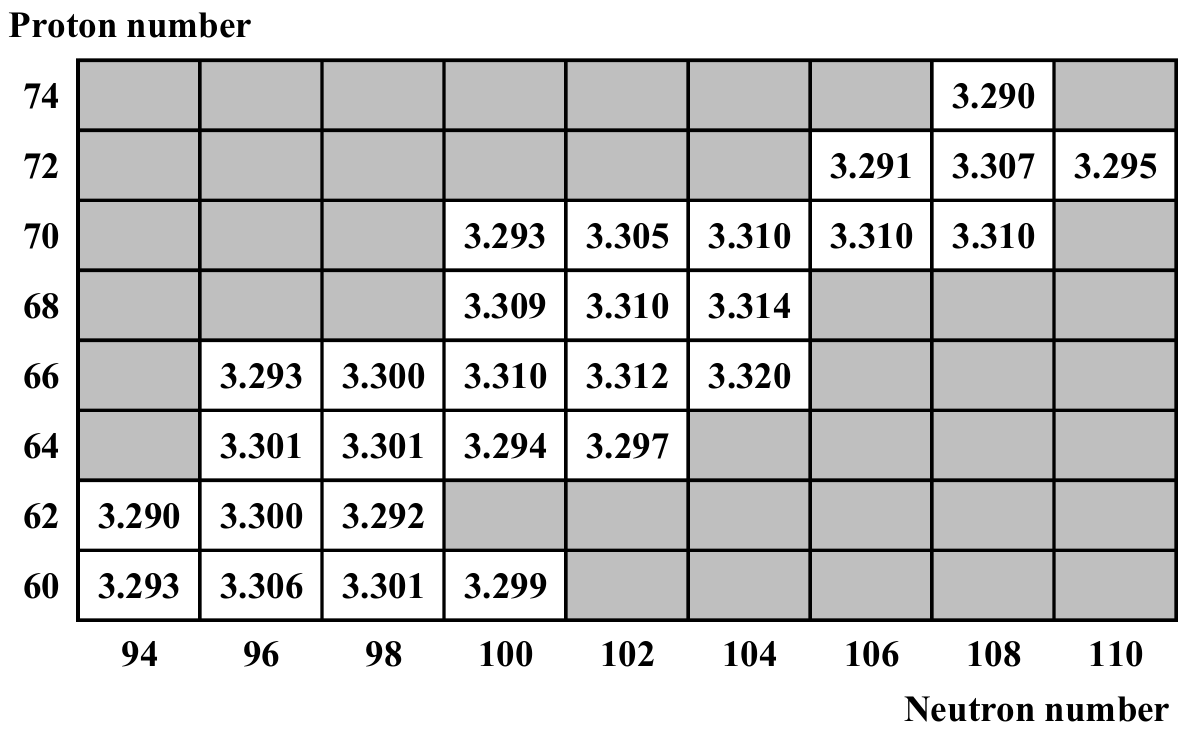}
    \caption{Ratio of the first $4^+$ over $2^+$ energies of the sample nuclei considered in this study.}
    \label{fig:R42 of sample nuclei}
\end{figure}

This is why we have chosen to use the Skyrme SIII parametrisation \cite{SIII} to describe the ground-state 
of some deformed nuclei belonging (or close to) the rare earth region. 
For the elements comprised in the interval $50 \le Z \le 82$, 
there are 28 isotopes (ranging from Neodymium to Tungsten) where the excitation energies of the first $4^+$
and $2^+$ states have a ratio $R_{4/2}$ which is larger than or equal to $3.290$ (data taken from Ref.~\cite{NNDC}). 
These nuclei are presented in Figure~\ref{fig:R42 of sample nuclei} together with the corresponding values of the $R_{4/2}$ energy ratio.

The average pair condensation energies $\tilde{E}_q^{pc}$ 
to be adjusted must be associated with the sp spectra corresponding to 
the calculated ground-state wavefunction which depends on the
definition of both the particle-hole potential and the pairing treatment in use. 
One problem arises \textit{a priori} then, since the reference values of $\tilde{E}_q^{pc}$
are determined in the seniority framework but are to be compared with those of a self-consistent solution of a calculation
performed in a residual pairing interaction framework. 
Between the two solutions there might exist some mean-field differences, 
such as a shift of the equilibrium deformation, resulting in different sp spectra.

This again suggests the pattern of an iterative approach. 
Starting from some intensity parameter values $W_q$,
one defines some ground-state spectra which are used to determine
the adequate seniority pair condensation energies $\tilde{E}_q^{pc} (sen)$.
Adjusting the parameters $W_q$ to yield values of $\tilde{E}_q^{pc}$
equal to the $\tilde{E}_q^{pc} (sen)$, one defines a new Hamiltonian giving rise 
to new sp spectra which are employed to determine new average
pair condensation energies and so on.

As illustrated in Appendix~\ref{Appendix:Volume delta}, this iterative process, which we have followed in our
present calculations, appears to be fastly convergent, upon using the knowledge of
values, e.g., from previous studies within a systematic investigation in a given region.
Thus in many cases, it appears that such iterations bring rather insignificant corrections.

On Table~\ref{tab: moi results} the results of four calculations are presented.

\begin{table*}
    \begin{ruledtabular}
        \caption{The moment of inertia obtained from individually fitted seniority strength $\mathcal{\gimel}^{(G)}$,
            from averaged seniority strengths $\mathcal{\gimel}^{(\tilde{G},ave)}$,
            from individually fitted delta force  $\mathcal{\gimel}^{W)}$ and 
            from the averaged delta strengths  $\mathcal{\gimel}^{(W,ave)}$
            are compared with the experimental ones  $\mathcal{\gimel}^{(exp)}$ derived from the first $2^+$ energies.}
        \label{tab: moi results}
        \begin{tabular}{*{12}c}
Element	&	Z	&	A	&	$\mathcal{\gimel}^{(exp)}$	&	$G_n$	&	$G_p$	&	$\mathcal{\gimel}^{(G)}$	&	
    $\mathcal{\gimel}^{(G,ave)}$	&	$W_n$	&	$W_p$	&	$\mathcal{\gimel}^{(W)}$	&	$\mathcal{\gimel}^{(W,ave)}$	\\
    \hline
Nd	&	60	&	154	&	42.373	&	19.687	&	18.260	&	45.459	&	45.368	&	-360	&	-385	&	44.982	&	43.586	\\
Nd	&	60	&	156	&	44.643	&	19.563	&	18.246	&	44.659	&	44.133	&	-363	&	-389	&	43.276	&	43.143	\\
Nd	&	60	&	158	&	45.524	&	19.405	&	18.299	&	44.613	&	43.760	&	-359	&	-395	&	43.403	&	43.899	\\
Nd	&	60	&	160	&	46.012	&	19.252	&	18.358	&	41.487	&	40.610	&	-359	&	-400	&	40.753	&	41.752	\\
Sm	&	62	&	156	&	39.531	&	19.814	&	18.510	&	38.171	&	39.256	&	-359	&	-406	&	37.076	&	38.412	\\
Sm	&	62	&	158	&	41.209	&	19.696	&	18.401	&	38.729	&	39.292	&	-360	&	-402	&	37.716	&	38.799	\\
Sm	&	62	&	160	&	42.373	&	19.549	&	18.333	&	41.832	&	41.809	&	-355	&	-400	&	41.257	&	41.532	\\
Gd	&	64	&	160	&	39.860	&	19.779	&	18.450	&	37.101	&	37.923	&	-359	&	-403	&	36.076	&	37.150	\\
Gd	&	64	&	162	&	41.899	&	19.645	&	18.344	&	41.624	&	41.984	&	-353	&	-400	&	41.117	&	40.928	\\
Gd	&	64	&	164	&	40.944	&	19.520	&	18.230	&	41.478	&	41.051	&	-353	&	-395	&	41.758	&	41.246	\\
Gd	&	64	&	166	&	42.857	&	19.400	&	18.159	&	44.437	&	43.164	&	-354	&	-393	&	45.297	&	44.741	\\
Dy	&	66	&	162	&	37.193	&	19.831	&	18.423	&	35.723	&	36.619	&	-359	&	-403	&	34.573	&	35.539	\\
Dy	&	66	&	164	&	40.876	&	19.722	&	18.360	&	40.168	&	40.904	&	-352	&	-400	&	39.598	&	39.241	\\
Dy	&	66	&	166	&	39.171	&	19.618	&	18.312	&	39.447	&	39.515	&	-352	&	-400	&	39.389	&	39.239	\\
Dy	&	66	&	168	&	40.021	&	19.501	&	18.258	&	42.015	&	41.582	&	-353	&	-400	&	42.320	&	42.070	\\
Dy	&	66	&	170	&	41.976	&	19.443	&	18.207	&	37.405	&	36.733	&	-361	&	-400	&	36.698	&	37.804	\\
Er	&	68	&	168	&	37.592	&	19.679	&	18.392	&	37.018	&	37.415	&	-350	&	-405	&	36.954	&	36.586	\\
Er	&	68	&	170	&	38.173	&	19.578	&	18.281	&	40.000	&	39.989	&	-350	&	-402	&	40.193	&	39.258	\\
Er	&	68	&	172	&	38.961	&	19.524	&	18.182	&	35.889	&	35.526	&	-358	&	-400	&	35.241	&	35.898	\\
Yb	&	70	&	170	&	35.606	&	19.719	&	18.414	&	36.155	&	36.711	&	-352	&	-405	&	35.796	&	35.635	\\
Yb	&	70	&	172	&	38.099	&	19.638	&	18.251	&	38.747	&	38.908	&	-350	&	-397	&	38.600	&	37.453	\\
Yb	&	70	&	174	&	39.231	&	19.588	&	18.083	&	36.325	&	35.722	&	-357	&	-390	&	35.713	&	35.427	\\
Yb	&	70	&	176	&	36.525	&	19.500	&	17.961	&	35.819	&	34.580	&	-356	&	-386	&	35.702	&	34.877	\\
Yb	&	70	&	178	&	35.714	&	19.390	&	17.858	&	38.632	&	36.536	&	-350	&	-382	&	39.457	&	37.206	\\
%Hf	&	72	&	176	&	33.956	&	19.644	&	18.147	&	32.444	&		&	-359	&	-391	&	31.682	&	32.035	\\
Hf	&	72	&	178	&	32.196	&	19.552	&	18.005	&	32.656	&	32.041	&	-355	&	-387	&	32.503	&	32.001	\\
Hf	&	72	&	180	&	32.146	&	19.473	&	17.902	&	34.472	&	33.449	&	-352	&	-388	&	34.637	&	33.718	\\
Hf	&	72	&	182	&	30.678	&	19.404	&	17.811	&	32.047	&	30.829	&	-357	&	-388	&	31.588	&	31.359	\\
%W	&	74	&	180	&	28.968	&	19.573	&	18.133	&	29.344	&		&	-355	&	-393	&	29.180	&	29.080	\\
W	&	74	&	182	&	29.968	&	19.502	&	18.001	&	30.949	&	30.476	&	-350	&	-386	&	31.513	&	30.354	\\
            \hline

        \end{tabular}
    \end{ruledtabular}
\end{table*}

The first set of results corresponds to seniority force pairing descriptions which
are a repetition of what has been performed in Ref.~\cite{Hafiza2019}. 
However, the present results differ slightly from them due to their more consistent (iterative) 
determination of the pairing matrix elements 
and also by dealing with a slightly larger sample (28 nuclei instead of 23). 
The average matrix elements are presented in terms of a 
nucleus-independent-fashion intensity parameter $G_q$
according to the parametrisation of Ref.~\cite{Bonche_pairing} taking into account the
dependence of the matrix elements on the considered nucleon number $N_q$ as
\begin{equation}
    V_q = \frac{G_q}{11 + N_q}.
\end{equation}

%The first column of MoI results concerns
The $\gimel^{(G)}$ MoI values are calculated with 
$G_q$ values obtained within an individual determination (that is nucleus per nucleus) 
process according to the iterative approach described above. 
They imply that the sp spectra correspond
to the equilibrium deformation obtained in a self-consistent HFBCS calculation 
using the resulting seniority force. 
The corresponding moments of inertia are compared with those deduced from the
experimental energies of the first $2^+$ states. 
The r.m.s deviation with the data of $\gimel^{(G)}$ moments is $2.05 \: \hbar^2 \mbox{MeV}^{-1}$.
The distribution of the $G_q$ values is defined by the following mean values $\tilde{G}_q$
and standard deviations $\sigma^q_G$:
\begin{flalign}
%    \item[] $\tilde{G}_n = 19.57$ MeV, \hspace{0.2cm} $\sigma^n_G = 0.13$ MeV
%    \item[] $\tilde{G}_p = 18.22$ MeV, \hspace{0.2cm} $\sigma^p_G = 0.20$ MeV.
    \tilde{G}_n = 19.57 \: \mbox{MeV},& \;\;\;\;\; \sigma^n_G = 0.13 \: \mbox{MeV}   \notag \\
    \tilde{G}_p = 18.22 \: \mbox{MeV},& \;\;\;\;\; \sigma^p_G = 0.20 \: \mbox{MeV}.
\end{flalign}

The second column of MoI results labeled $\gimel^{(G,ave)}$  concern
the moments of inertia obtained in self-consistent
HFBCS calculations within the seniority force framework using the above displayed averaged 
$\tilde{G}_n$ and $\tilde{G}_p$ values. 
The r.m.s. difference with the data is $2.04 \: \hbar^2 \mbox{MeV}^{-1}$.

The next set of results corresponds to
calculations performed with a volume delta interaction.

The third column of MoI results labeled $\gimel^{(W)}$  concerns
an individual determination of the parameters $W_q$.
The distribution of these values is defined by the following mean values 
$\tilde{W}_q$ and standard deviations $\sigma^q_W$:
\begin{flalign}
%    \item[] $\tilde{W}_n = -355$ MeV, \hspace{0.2cm} $\sigma^n_W = 4$ MeV
%    \item[] $\tilde{W}_p = -396$ MeV, \hspace{0.2cm} $\sigma^p_W = 7$ MeV.
    \tilde{W}_n = -355 \: \mbox{MeV},& \;\;\;\;\; \sigma^n_W = 4 \: \mbox{MeV}   \notag \\
    \tilde{W}_p = -396 \: \mbox{MeV},& \;\;\;\;\; \sigma^p_W = 7 \: \mbox{MeV}.
    \label{eq: standard deviation W_q}
\end{flalign}

As in the previous case, we compare the calculated and experimentally deduced moments of
inertia. The r.m.s deviation with the data is $2.53 \: \hbar^2 \mbox{MeV}^{-1}$.

Finally, the fourth column of MoI results labeled $\gimel^{(\tilde{W})}$
concerns calculations using the same
delta interactions with the above average parameters $\tilde{W}_q$
for all the nuclei in the sample. 
The corresponding r.m.s deviation with the data is now $1.91 \: \hbar^2 \mbox{MeV}^{-1}$.

What can we conclude from these tests of our method when comparing calculated moments
of inertia with those deduced from experimental first $2^+$ state energies? 
First of all, in the seniority case, using individually-fitted 
or averaged matrix elements does not induce significant differences. 
This is not so, on the contrary, in the delta interaction case where the average
interaction provides significantly better results than the individually-fitted ones. 
It could be that any possible inadequacy in the calculated spectra for some nuclei is washed out by the averaging
with respect either to the sp energies around the Fermi energy in the seniority approximation case 
or to the number of nucleons in the delta interaction case. 
In the latter case, this could explain why the averaging of the interactions yields an improvement of about 20\% in the
reproduction of the moments of inertia, 
making it to reach the phenomenological quality of the seniority approach for this particular nuclear property.

\section{Conclusion
\label{sec:Section V}}

We have proposed a method to adjust the parameters of residual (neutron-neutron and proton-proton) pairing volume delta interactions. 
Its phenomenological value has been assessed by its ability to reproduce the
moments of inertia of all the well and rigidly deformed isotopes of the
elements whose proton numbers $Z$ are such that $50 \le Z \le 82$. 
It has been demonstrated that this approach leads to a reproduction of the same
quality as direct fits of these moments
within a seniority force approach.

The pair condensation energy averaged \textit{\`{a} la} Strutinsky (which is essentially
contingent upon the value of the sp energy density at the Fermi energy) has 
appeared to be a key index to determine the intensity of pairing correlations.

We have determined the parameters of these volume delta interactions individually, that is nucleus per nucleus. 
Yet, we have shown that averaging them over a given nuclear region provided a better overall reproduction of
the moments of inertia. 
It may be arguably considered that the mere concept of universal or even regional residual interactions is by definition
irrelevant, since they are what remains of a supposedly universal (or regional) interaction by
subtraction of 
a mean-field which is by essence solution-dependent. 
Furthermore, the situation is somewhat obscured by the consideration in current phenomenological mean-field calculations of
density-dependent effective interactions (as, e.g., in the Skyrme or Gogny cases) 
or even more so when they are performed within the so-called Energy Density Functional formalism.

Besides there are more practical reasons why they should to some extent vary, from one solution to another. 
The first one is due to the approximation retained to treat pairing correlations which may vary
from one nuclear state to another. 
For instance, as we have already mentioned, the BCS approximation happens to be rather poor in weak
pairing regimes \cite{Zheng_1992}. 
A local fit would tend to correct unduly for that. 
The second reason is related to the necessity of defining an energy
cut-off in the sp energies or as in our case, limiting to a given sp energy
interval the sp states participating in the variational BCS approach. 
The parameter of the cut-off definition (the size $2\omega$ of the energy interval in
our case) then becomes a parameter of the interaction which cannot easily
be adjusted to the passage from solution to another (as, e.g., to cope with the overall nuclear $A^{-1/3}$
energy scaling or a possible deformation dependence). 
In this respect, let us remark that, in our case, the difference in the intensity parameters $W_q$
between protons and neutrons are not only due to the Coulomb anti-pairing 
but mostly to the fact that we have chosen the same cut-off parameter $2 \omega$
for neutrons and protons.

One of the motivations to go beyond the seniority approach for the pairing
treatment, apart from accounting for a sp state-dependence of the pairing gaps, is
to allow for an extrapolation as safe as possible, from a description limited
to some specific states of specific nuclei. 
In order to be able to do that, we have no other choice than defining in a more or less extended region of the nuclear chart, 
a single residual interaction whose relevance and predictive power is, 
as always in any phenomenological approach, to be validated by
its ability to reproduce some existing data other than those on which it is grounded.

The present paper has limited its scope to the choice of a particular
analytical form of the residual interaction and to the BCS approximation
for the pairing treatment. 
In so far as what counts to quantitatively define
the intensity of pairing correlations 
(producing a given amount of diffusion of pairs near the Fermi surface) 
is the average pair condensation energy, 
one is allowed to expect that what has been achieved here could be successfully 
extended to other pairing interactions and possibly to other theoretical pairing approaches, 
provided that they rely on some relevant sp spectra. 
Work in these directions is under way.

\begin{acknowledgments}
M.H.K acknowleges Universiti Teknologi Malaysia for its
UTM Fundamental Research grant (grant number Q.J130000.3854.23H49).
The same author also extend appreciation to Dr. Razif Razali and UTM Digital for the use
of the high performance computing facility where some test calculations were performed.
\end{acknowledgments}

\appendix
\section{Some numerical details about the HF-BCS calculations
\label{Appendix:HFBCS numerical details}}
In our calculations, axial symmetry is imposed. 
The eigen-solutions of the HF Hamiltonian are obtained by expanding canonical basis states 
onto a truncated basis of axially-symmetrical harmonic oscillator eigen-states. 
The latter depend on two parameters $\omega_z$ and $\omega_\perp$, 
corresponding to the oscillator frequencies on the symmetry axis $O_z$ and in the perpendicular plane. 
An equivalent spherical oscillator frequency $\omega_0$ is defined by $\omega_0^3 = \omega_z \omega_\perp^2$. 
As in Reference~\cite{NPA203}, we use two basis parameters instead: 
a size parameter $b = \sqrt{m \omega_0/\hbar}$ where $m$ is a charge averaged nucleonic mass 
and a deformation parameter $q = \omega_\perp/\omega_z$. 
A deformation-dependent truncation prescription (see \cite{NPA203}) is defined by a parameter $N_0$ 
corresponding to the maximum total number of oscillator quanta at sphericity. 
In our calculations $N_0 = 16$. 
The truncation entails a dependence of the eigen-solutions upon the $(b,q)$ basis parameters at a given $N_0$ value. 
These parameters are determined so as to minimize the total energy. 
The values of the basis parameters in use are given in Table~\ref{tab: basis parameters}. 
Calculations of various matrix elements and energies are performed within a Gauss-Hermite and Gauss-Laguerre 
(in the cylindrical variables $z$ and $\rho$, respectively) approximation with $N_z = 30$ and $N_\perp = 15$ mesh points.

\begin{table}
    \begin{ruledtabular}
        \caption{The values of basis parameters $b$ and $q$ used in the calculations
                for a basis size of $N_0 = 16$.}
        \label{tab: basis parameters}
        \begin{tabular}{*{5}c}
        Element&    $Z$&    $A$&  $b$&  $q$ \\
        \hline
Nd	&	60	&	154	&	0.490	&	1.240	\\
Nd	&	60	&	156	&	0.490	&	1.240	\\
Nd	&	60	&	158	&	0.490	&	1.240	\\
Nd	&	60	&	160	&	0.490	&	1.240	\\
Sm	&	62	&	156	&	0.493	&	1.239	\\
Sm	&	62	&	158	&	0.491	&	1.239	\\
Sm	&	62	&	160	&	0.490	&	1.239	\\
Gd	&	64	&	160	&	0.490	&	1.189	\\
Gd	&	64	&	162	&	0.488	&	1.189	\\
Gd	&	64	&	164	&	0.487	&	1.189	\\
Gd	&	64	&	166	&	0.485	&	1.189	\\
Dy	&	66	&	162	&	0.488	&	1.189	\\
Dy	&	66	&	164	&	0.487	&	1.189	\\
Dy	&	66	&	166	&	0.485	&	1.189	\\
Dy	&	66	&	168	&	0.484	&	1.189	\\
Dy	&	66	&	170	&	0.482	&	1.189	\\
Er	&	68	&	168	&	0.484	&	1.189	\\
Er	&	68	&	170	&	0.482	&	1.189	\\
Er	&	68	&	172	&	0.480	&	1.189	\\
Yb	&	70	&	170	&	0.482	&	1.189	\\
Yb	&	70	&	172	&	0.481	&	1.189	\\
Yb	&	70	&	174	&	0.479	&	1.189	\\
Yb	&	70	&	176	&	0.477	&	1.189	\\
Yb	&	70	&	178	&	0.476	&	1.189	\\
Hf	&	72	&	176	&	0.477	&	1.189	\\
Hf	&	72	&	178	&	0.476	&	1.189	\\
Hf	&	72	&	180	&	0.474	&	1.184	\\
Hf	&	72	&	182	&	0.471	&	1.189	\\
W	&	74	&	180	&	0.476	&	1.140	\\
W	&	74	&	182	&	0.476	&	1.140	\\
        \end{tabular}
    \end{ruledtabular}
\end{table}

\section{Convergence of the two iterative processes
\label{Appendix:convergence of iterative process}}

Let us specify the notation of the various pairing gaps appearing in what follows.

First, we consider pairing gaps corresponding to the quantal solutions of the HFBCS equation.
In the seniority case, they are noted as $\Delta_q^{(m)}$, $q$ standing for neutrons and protons and 
$m$ referring to the $m^{th}$ step of the iterative process described below.

In calculations making use of  a residual interaction, they are noted as $\Delta_{q,i}^{(m)}$
where $i$ corresponds to the label of a given Cooper pair.

Second, we consider the average pairing gaps issued from the phenomenological approach of M\"{o}ller and Nix \cite{Moller1992}.
They are noted generically as $\Delta_q^{MN}$.
However, as discussed below, in the proton case they are corrected with a 
corrective factor which varies along the iterative steps.
This is why to differentiate them from the $\Delta_p^{MN}$ of Ref.~\cite{Moller1992},
we will note them $\Delta_p^{MN (m)}$ when defined at the $m^{th}$ step of the iterative process.

\subsection{Seniority pairing treatment
\label{Appendix:seniority treatment}}

The absolute value of the pair condensation energy is defined as
\begin{equation}
    E_q^{pc} = \sum_i k_{q,i} \Delta_{q,i}
\end{equation}
where the summation runs on Cooper pairs.

The corresponding average pair condensation takes a particular form in the seniority case since there
\begin{equation}
    \forall i, \Delta_{q,i} = \tilde{\Delta}_{q,i} = \Delta_q \:\:\: \mbox{and} \:\:\: \sum_i \tilde{k}_i = \sum_i k_i.
\end{equation}

Consequently, with 
\begin{equation}
    \Delta_q = - V_q \sum_i k_i
\end{equation}
one has 
\begin{equation}
    \tilde{E}_q^{pc} (sen) = E_q^{pc} (sen) = \frac{\Delta^2}{V_q}.
\end{equation}

Starting with an ansatz $V_q^{(0)}$
for the average matrix elements, we perform a self-consistent seniority HFBCS-S calculation providing initial 
sp spectra $\{\epsilon_q\}^{(0)}$ and pairing gaps $\Delta_q^{(0)}$.

A difficulty arises with the uniform-gap method in the proton case due to 
the use in our HFBCS code of the Slater approximation \cite{Slater} for the treatment of the Coulomb exchange terms 
(see Subsection II.D of Ref.~\cite{Koh_Philippe_2024}). 

To correct this, using the pair condensation energy for protons 
calculated in this first step of the iterative process $E_p^{pc (0)}$, 
we can renormalize the M\"{o}ller-Nix phenomenological average
proton gaps $\Delta_p^{MN}$ with a multiplicative factor evaluated approximately (see \cite{Koh_Philippe_2024}) as
\begin{equation}
    R_p = 0.0181 \: E_p^{pc} (sen) + 0.781
\end{equation}
(where $E_p^{pc (0)}$ is given in MeV). 
This provides a corrected M\"{o}ller-Nix gap
\begin{equation}
    \Delta_p^{MN (0)} = R_p \: \Delta_p^{MN}
\end{equation}
to be used in Eq. (\ref{eq: uniform gap equation}).

The solutions $V_q^{(1)}$ of the uniform-gap method (see Equation~\ref{eq: uniform gap equation})
are then implemented in a second run of the self-consistent HFBCS-S approach. 
This yields new sp spectra $\{\epsilon_q\}^{(1)}$ and gaps $\Delta_q^{(1)}$
from which we determine as 
above described new average matrix elements $V_q^{(2)}$
and so on.
The convergence is considered to have been reached when the results for the
$G_q$ are stable up to 10 keV.

The fast character of the convergence is illustrated on Table~\ref{tab: convergence of iterative process seniority force} 
for 4 nuclei ($^{156}$Nd, $^{164}$Gd, $^{172}$Er, $^{180}$Hf).

\begin{table}
    \begin{ruledtabular}
        \caption{The neutron and proton strengths of the estimated seniority force at specific iterations.
                The starting strengths of $G_q = 19$ MeV is used.
                Convergence is achieved after 3 iterations.
                The iteration number is referred to as $m$.}
        \label{tab: convergence of iterative process seniority force}
        \begin{tabular}{*{6}c}
            Nucleus& $Z$& $A$& $m$& $G_n$& $G_p$    \\
            \hline
            \multirow{4}{*}{Nd}&    \multirow{4}{*}{60}&     \multirow{4}{*}{156}&    0&  19.00&   19.00   \\ 
            &&& 1&  19.56&   18.25    \\
            &&& 2&  19.58&   18.10    \\
            &&& 3&  19.58&   18.07    \\
            \hline
            \multirow{4}{*}{Gd}&    \multirow{4}{*}{64}&     \multirow{4}{*}{164}&    0&  19.00&   19.00   \\ 
            &&& 1&  19.52&   18.23    \\
            &&& 2&  19.54&   18.07    \\
            &&& 3&  19.55&   18.04    \\
            \hline
            \multirow{4}{*}{Er}&    \multirow{4}{*}{68}&     \multirow{4}{*}{172}&    0&  19.00&   19.00   \\
            &&& 1&  19.52&   18.18    \\
            &&& 2&  19.55&   18.00    \\
            &&& 3&  19.55&   17.97    \\
            \hline
            \multirow{4}{*}{Hf}&    \multirow{4}{*}{72}&     \multirow{4}{*}{180}&    0&  19.00&   19.00   \\
            &&& 1&  19.47&   17.90    \\
            &&& 2&  19.49&   17.66    \\
            &&& 3&  19.49&   17.61    \\
        \end{tabular}
    \end{ruledtabular}
\end{table}

\subsection{Volume delta interaction pairing treatment
\label{Appendix:Volume delta}}
Starting with an ansatz $W_q^{(0)}$
for the delta interaction intensities, 
we perform a self-consistent seniority HFBCS calculation providing sp spectra $\{\epsilon_q^i\}^{(0)}$
as well as sets of $\{k_p^i\}^{(0)}$ factors and pairing gaps $\{\Delta_p^i\}^{(0)}$ for protons. 
From this we evaluate the average proton pair condensation energy $E_p^{pc (0)}$
associated with the HFBCS solution. 
As in Subsection~\ref{Appendix:seniority treatment},
we then are able to determine a corrected M\"{o}ller-Nix proton gap
\begin{equation}
    \Delta_p^{MN (0)} = R_p \: \Delta_p^{MN}
\end{equation}
with $R_p = 0.0181 \: E_p^{pc (0)} + 0.781$
(where $E_p^{pc (0)}$ is given in MeV).

Solving the uniform-gap equations using the spectra $\{\epsilon_q^i\}^{(0)}$
and the M\"{o}ller-Nix gaps ($\Delta_n^{MN},\Delta_p^{MN (0)} $),
we get the matrix elements $V_q^{(0)}$
and the seniority pairing gaps $\{\Delta_q\}^{(0)}$
from which we evaluate the average seniority
pair condensation energies
\begin{equation}
    E_q^{pc (0)} (sen) = \frac{{\Delta_q^{(0)}}^2}{V_q^{(0)}}.
\end{equation}

We then perform a series of HFBCS calculations varying the intensities
$V_q$ of the delta interaction so as to match the resulting pair
condensation energies averaged as above described with the average seniority pair
condensation energies $E_q^{pc (0)}$.

This provides new intensity parameters $W_q^{(1)}$
yielding sp spectra $\{\epsilon_q^i\}^{(1)}$,
$\{k_q^i\}^{(1)}$ factors and pairing gaps $\{\Delta_q^{(1)}\}$.
We then evaluate a new corrected M\"{o}ller-Nix proton gap $\{\Delta_p^{MN (1)}\}$.

Solving the uniform-gap equations using the spectra $\{\epsilon_q^i\}^{(1)}$
and the M\"{o}ller-Nix gaps $(\Delta_n^{MN},{\Delta_p^{MN}}^{(1)})$,
we get the matrix elements $V_q^{(1)}$, 
the seniority pairing gaps $\Delta_q^{(1)}$
and the proton average seniority pair condensation energies $E_p^{pc (1)} = \frac{{\Delta_q^{(1)}}^2}{V_q^{(1)}}$.

We continue the iterative process performing a new series of HFBCS
calculations to provide new intensities $W_q^{(2)}$
up to a convergence considered to have been reached when the values of the
intensities $W_q$ are stable up to  $10$ keV.

As in the Subsection~\ref{Appendix:seniority treatment}, the fast character of the convergence is
illustrated on Table~\ref{tab: convergence of iterative process delta force} 
for the same 4 nuclei ($^{156}$Nd, $^{164}$Gd, $^{172}$Er, $^{180}$Hf).
We found that most calculations converged at the second iterations.

It has also been checked that starting with a different initial $W_q$,
we obtained estimated intensities that differ by a couple of MeV.
More specifically for the four considered nuclei in Table~\ref{tab: convergence of iterative process delta force},
using $W_n^{(0)} = -300$ MeV and $W_p^{(0)} = -350$ MeV as starting values, 
we found a difference at most of about 1.5 MeV for $W_n$ and about $2.0$ MeV for $W_q$.
%which are well within the standard deviation $\sigma_q$ in Eq.~(\ref{eq: standard deviation W_q}).
%These differences are poorly affecting the results when 
Thus, even for such large differences in initial values of $W_q$,
the resulting differences in interaction intensities 
are poorly affecting the final values of the interaction parameters 
in view of the variance of their distribution as displayed in Eq.~(\ref{eq: standard deviation W_q}).
Indeed, the final value of the strength $\widetilde W_q$ 
then results from an average over the converged values of $W_q$ obtained for several nuclei.

\begin{table}
    \begin{ruledtabular}
        \caption{The estimated delta strengths $W_q$ in MeV with convergence achieved even at the first iteration.
        The iteration number is referred to as $m$.}
        \label{tab: convergence of iterative process delta force}
        \begin{tabular}{*{6}c}
            Nucleus& $Z$& $A$& $m$& $W_n$& $W_p$    \\
            \hline
            \multirow{3}{*}{Nd}&    \multirow{3}{*}{60}&     \multirow{3}{*}{156}&    0&  -350.00&   -400.00   \\ 
            &&& 1&  -362.59&   -389.20    \\
            &&& 2&  -362.59&   -389.20    \\
            \hline
            \multirow{3}{*}{Gd}&    \multirow{3}{*}{64}&     \multirow{3}{*}{164}&    0&  -350.00&   -400.00   \\ 
            &&& 1&  -352.94&   -394.52    \\
            &&& 2&  -352.94&   -394.52    \\
            \hline
            \multirow{3}{*}{Er}&    \multirow{3}{*}{68}&     \multirow{3}{*}{172}&    0&  -350.00&   -400.00   \\
            &&& 1&  -358.45&   -400.00    \\
            &&& 2&  -358.45&   -400.00    \\
            \hline
            \multirow{3}{*}{Hf}&    \multirow{3}{*}{72}&     \multirow{3}{*}{180}&    0&  -350.00&   -400.00   \\
            &&& 1&  -352.14&   -388.31    \\
            &&& 2&  -352.14&   -388.31    \\
        \end{tabular}
    \end{ruledtabular}
\end{table}

\bibliography{apssamp}% Produces the bibliography via BibTeX.

\end{document}